\documentclass[prd,superscriptaddress,10pt,twocolumn,preprintnumbers]{revtex4}

\usepackage[latin1]{inputenc}
\usepackage{graphicx}
\usepackage[english]{babel}
\usepackage[hidelinks]{hyperref}
\usepackage{amsmath}
\usepackage{amssymb}
\usepackage{amsfonts}
\usepackage{psfrag}
\usepackage{colortbl} 
\usepackage[dvipsnames]{xcolor} 
\usepackage{pifont} 
\usepackage{amsmath} 

\definecolor{MyBlue}{rgb}{0.001,0.001,0.99}

\hypersetup{
colorlinks=true,
citecolor=MyBlue,
linkcolor=MyBlue,
urlcolor=MyBlue
}

\begin{document}
\def\cL{{\cal L}}
%\def\be{\begin{equation}}
%\def\ee{\end{equation}}
%\def\bea{\begin{eqnarray}}
%\def\eea{\end{eqnarray}}
%\def\beq{\begin{eqnarray}}
%\def\eeq{\end{eqnarray}}
%\def\tr{{\rm tr}\, }
%\def\nn{\nonumber \\}
%\def\e{{\rm e}}

%\title{Axially symmetric solutions in $f(R)$-gravity}
%
%
%\author{Salvatore Capozziello$^1$, Mariafelicia De Laurentis$^1$, Arturo Stabile$^2$}
%\affiliation{\it $^1$Dipartimento di Scienze Fisiche, Università
%di Napoli {}``Federico II'', INFN Sez. di Napoli, Compl. Univ. di
%Monte S. Angelo, Edificio G, Via Cinthia, I-80126, Napoli, Italy\\ $^2$ Dipartimento di Ingegneria,
%Università del Sannio,
%Benevento, C.so Garibaldi
%107, I-80125 Benevento, Italy}
%\date{\today

%\begin{document}

\def\bef{\begin{figure}}
\def\eef{\end{figure}}
\newcommand{\ans}{ansatz }
\newcommand{\eeqn}{\end{eqnarray}}
\newcommand{\bd}{\begin{displaymath}}
\newcommand{\ed}{\end{displaymath}}
\newcommand{\mat}[4]{\left(\begin{array}{cc}{#1}&{#2}\\{#3}&{#4}
\end{array}\right)}
\newcommand{\matr}[9]{\left(\begin{array}{ccc}{#1}&{#2}&{#3}\\
{#4}&{#5}&{#6}\\{#7}&{#8}&{#9}\end{array}\right)}
\newcommand{\matrr}[6]{\left(\begin{array}{cc}{#1}&{#2}\\
{#3}&{#4}\\{#5}&{#6}\end{array}\right)}
\newcommand{\cvb}[3]{#1^{#2}_{#3}}
\def\lsim{\raise0.3ex\hbox{$\;<$\kern-0.75em\raise-1.1ex
e\hbox{$\sim\;$}}}
\def\gsim{\raise0.3ex\hbox{$\;>$\kern-0.75em\raise-1.1ex
\hbox{$\sim\;$}}}
\def\abs#1{\left| #1\right|}
\def\simlt{\mathrel{\lower2.5pt\vbox{\lineskip=0pt\baselineskip=0pt
           \hbox{$<$}\hbox{$\sim$}}}}
\def\simgt{\mathrel{\lower2.5pt\vbox{\lineskip=0pt\baselineskip=0pt
           \hbox{$>$}\hbox{$\sim$}}}}
\def\unity{{\hbox{1\kern-.8mm l}}}
\newcommand{\eps}{\varepsilon}
\def\ep{\epsilon}
\def\ga{\gamma}
\def\Ga{\Gamma}
\def\om{\omega}
\def\omp{{\omega^\prime}}
\def\Om{\Omega}
\def\la{\lambda}
\def\La{\Lambda}
\def\al{\alpha}
\newcommand{\ov}{\overline}
\renewcommand{\to}{\rightarrow}
\renewcommand{\vec}[1]{\mathbf{#1}}
\newcommand{\vect}[1]{\mbox{\boldmath$#1$}}
\def\tm{{\widetilde{m}}}
\def\mcirc{{\stackrel{o}{m}}}
\newcommand{\Dm}{\Delta m}
\newcommand{\dm}{\varepsilon}
\newcommand{\tanb}{\tan\beta}
\newcommand{\nbar}{\tilde{n}}
\newcommand\PM[1]{\begin{pmatrix}#1\end{pmatrix}}
\newcommand{\up}{\uparrow}
\newcommand{\down}{\downarrow}
\def\omE{\omega_{\rm Ter}}
%
%%%%%%%%%%     mauri    %%%%%%%%%%%%%%%%%%%%%%%%%%%%%%%%%

\newcommand{\Dsusy}{{susy \hspace{-9.4pt} \slash}\;}
\newcommand{\DCP}{{CP \hspace{-7.4pt} \slash}\;}
\newcommand{\mc}{\mathcal}
\newcommand{\gr}{\mathbf}
\renewcommand{\to}{\rightarrow}
\newcommand{\gtc}{\mathfrak}
\newcommand{\wh}{\widehat}
\newcommand{\br}{\langle}
\newcommand{\kt}{\rangle}

%%%%%%%%%%%%%%%%%%%%%%%%%%%%%%%%%%%%%%%%%%%%%%%%%%%%%%%%%%

% barbara Ricci  %definizione di minore e maggiore simile
\def\lsim{\mathrel{\mathop  {\hbox{\lower0.5ex\hbox{$\sim$}
\kern-0.8em\lower-0.7ex\hbox{$<$}}}}}
\def\gsim{\mathrel{\mathop  {\hbox{\lower0.5ex\hbox{$\sim$}
\kern-0.8em\lower-0.7ex\hbox{$>$}}}}}
%%%%%%%%%%%%%%%%%%%%%%%%%%%%%%%%%%

\def\bla{{\mathbf \la}}
\def\bmu{\mathbf m }
\def\by{{\mathbf y}}
\def\bmu{\mbox{\boldmath $\mu$} }
\def\bsig{\mbox{\boldmath $\sigma$} }
\def\bunity{{\mathbf 1}}

\def\be{\begin{equation}}
\def\ee{\end{equation}}
\def\bea{\begin{eqnarray}}
\def\eea{\end{eqnarray}}
\def\beq{\begin{eqnarray}}
\def\eeq{\end{eqnarray}}
\def\bas{\begin{subequations}\begin{eqnarray}}
\def\eas{\end{eqnarray}\end{subequations}}
\def\nn{\nonumber}
\def\sgn{\mathrm{sgn}}
\def\eps{\varepsilon}
\def\lp{\ell_\text{Pl}}
\def\Tr{\text{Tr}}
\def\tr{\text{tr}}
\def\la{\langle}
\def\ra{\rangle}
\def\de{\mathrm{d}}
\def\Pexp{\overrightarrow{\exp}}
\def\f{\frac}
\def\lb{\big\lbrace}
\def\rb{\big\rbrace}
\def\SU{\text{SU}}
\def\SO{\text{SO}}
\def\SP{\text{Spin}}
\def\SL{\text{SL}}
\def\su{\mathfrak{su}}
\def\so{\mathfrak{so}}
\def\sl{\mathfrak{sl}}
\def\g{{}^{\text{\tiny{($\gamma$)}}}\!}
\def\vp{\vphantom{-1}}
\def\openone{\mathds{1}}
\def\time{\,{\scriptstyle{\times}}\,}
\def\L{{\text{\tiny{L}}}}
\def\E{{\text{\tiny{E}}}}
\def\lc{\ell_\text{c}}
\def\DSU{\text{DSU}}
\newcommand{\cB}{{\mathbb B}}

\newcommand{\C}{{\mathbb C}}
\newcommand{\N}{{\mathbb N}}
\newcommand{\R}{{\mathbb R}}
\newcommand{\Z}{{\mathbb Z}}

\newcommand{\cY}{{\mathcal Y}}

\newcommand{\cE}{{\mathcal E}}

\newcommand{\cI}{{\mathcal I}}
\newcommand{\cJ}{{\mathcal J}}
\newcommand{\cK}{{\mathcal K}}
\newcommand{\cH}{{\mathcal H}}
\newcommand{\cM}{{\mathcal M}}
\newcommand{\cN}{{\mathcal N}}
\newcommand{\cQ}{{\mathcal Q}}
\newcommand{\cR}{{\mathcal R}}
\newcommand{\cO}{{\mathcal O}}
\newcommand{\cP}{{\mathcal P}}
\newcommand{\cT}{{\mathcal T}}
\newcommand{\cV}{{\mathcal V}}
\newcommand{\cD}{{\mathcal D}}
\newcommand{\cC}{{\mathcal C}}
\newcommand{\cS}{{\mathcal S}}
\newcommand{\cU}{{\mathcal U}}
\newcommand{\cZ}{{\mathcal Z}}

\newcommand{\Jib}{\color{red}}
\newcommand{\Etera}{\color{blue}}

\def\tgamma{\tilde{\gamma}}
\def\tc{\tilde{c}}
\def\tp{\tilde{p}}
\def\tv{\tilde{v}}
\def\tb{\tilde{b}}
\def\vj{\vec{\j}}
\def\veta{\vec{\eta}}
\def\bz{\bar{z}}
\def\vtau{\vec{\tau}}
\def\pp{\partial}
\def\rd{\textrm{d}}

\def\ka{\kappa}
\def\vphi{\varphi}
\def\eps{\epsilon}
\def\om{\omega}

\newcommand{\id}{\mathbb{I}}
\def\ot{\otimes}
\def\one{{\bf 1}}
\def\act{\triangleright}

\def\balpha{\bar{\alpha}}
\def\bbeta{\bar{\beta}}
\def\vJ{\vec{J}}
\def\vlambda{\vec{\lambda}}
\def\vepsilon{\vec{\epsilon}}
\def\vnu{\vec{\nu}}
\def\vtheta{\vec{\theta}}
\def\vtau{\vec{\tau}}
\def\tk{\tilde{k}}
\def\tj{\tilde{j}}
\def\la{\langle}
\def\ra{\rangle}

\newcommand{\Ji}{\color{red}}
\newcommand{\Et}{\color{blue}}
\newcommand{\y}{ { \boldsymbol{\rm y}} }

\newcommand{\Sp}{\mathrm{Sp}}
\newcommand{\ISU}{\mathrm{ISU}}
\newcommand{\Spin}{\mathrm{Spin}}

\renewcommand{\O}{\mathrm{O}}
\newcommand{\U}{\mathrm{U}}
\newcommand{\ISO}{\mathrm{ISO}}
\newcommand{\SB}{\mathrm{SB}}

\newcommand{\bes}{\begin{eqnarray}}
\newcommand{\ees}{\end{eqnarray}}

\renewcommand{\sp}{{\mathfrak{sp}}}
\newcommand{\spin}{{\mathfrak{spin}}}
\renewcommand{\sl}{{\mathfrak{sl}}}

\newcommand{\qq}{\,,\quad}
\newcommand{\tl}{\widetilde}
\def\nn{\nonumber}
\def\pp{\partial}
\newcommand{\w}{\wedge}

\def\ka{\kappa}
\def\vphi{\varphi}
\def\eps{\epsilon}
\def\om{\omega}

\def\ot{\otimes}
\def\one{{\bf 1}}
\def\act{\triangleright}

\def\vx{\vec{x}}
\def\vu{\vec{u}}
\def\vp{\vec{p}}
\def\vq{\vec{q}}
\def\vv{\vec{v}}
\def\vcJ{\vec{\cJ}}
\def\vJ{\vec{J}}
\def\vK{\vec{K}}
\def\vL{\vec{L}}
\def\vN{\vec{N}}
\def\vsigma{\vec{\sigma}}
\def\arr{\rightarrow}
\def\act{\,\triangleright\,}
\def\bz{\bar{z}}
\def\bZ{\bar{Z}}

\def\hn{\hat{n}}
\def\hu{\hat{u}}
\def\he{\hat{e}}
\def\hv{\hat{v}}
\def\hp{\hat{p}}
\def\hJ{\hat{J}}
\def\hcJ{\widehat{\cJ}}

\def\tT{\tilde{T}}
\def\tN{\tilde{N}}
\def\tcG{\widetilde{\cG}}
\def\tG{\widetilde{G}}
\def\tt{\tilde{t}}
\def\dd{\mathrm{d}}
\def\vX{\vec{X}}
\def\vcC{\vec{\cC}}
\def\tx{\tilde{x}}
\def\tp{\tilde{p}}
\def\tj{\tilde{j}}
\def\tk{\tilde{k}}
\def\wcH{\widetilde{\cH}}

\title{Hairy Schwarzschild-(A)dS black hole solutions in DHOST theories \\ beyond shift symmetry}

\author{{ J. Ben Achour}}\email{jibrilbenachour@gmail.com}
\affiliation{Center for Relativity and Gravitation, Beijing Normal University, Beijing 100875, China}

\author{H. Liu}
\email{liu.hongguang@cpt.univ-mrs.fr}
\affiliation{Aix Marseille Universit\'e, Universit\'e de Toulon, CNRS, CPT, Marseille, France}
\affiliation{Institut Denis Poisson, Universit\'e d'Orl\'eans, Universit\'e de Tours, CNRS,  37200 Tours, France}
\affiliation{Laboratoire AstroParticule et Cosmologie,  Universit\'e Paris Diderot Paris 7, CNRS, 75013 Paris, France}

%\date{\today}
\begin{abstract}
\noindent 
%Extensions of General Relativity (GR) face the challenging task to modify gravity on large scale while restoring GR predictions on scales in which the theory is experimentally confirmed, typically for the local tests within the solar system. In this regard, stealth black hole solutions provide an interesting path to discuss minimal hairy black holes backgrounds beyond GR. 
We derive the general conditions for a large family of shift symmetry breaking degenerate higher order scalar-tensor (DHOST) theories to admit stealth black hole solutions. Such black hole configurations correspond to vacuum solutions of General Relativity and admit a scalar hair which does not gravitate, revealing itself only at the perturbative level. 
%It was recently shown that the subclass of shift symmetry breaking Horndeski theories, consistent with the recent observational constraint $c_{\text{grav}}=c_{\text{light}}$, does not admit such stealth configuration when the scalar field is time-dependent. In this work, we extend this result and derive the conditions for a large class of DHOST theories to admit 
We focus our investigation on hairy Schwarzschild-(A)dS or pure Schwarzschild solutions, dressed with a linear time-dependent scalar hair, and assuming a constant kinetic term. We also discuss subclasses of this family which satisfy the observational constraint $c_{\text{grav}}=c_{\text{light}}$, as well as the recent constraint ensuring the absence of graviton decay. We provide at the end concrete examples of DHOST lagrangians satisfying our conditions. This work provides a first analysis of exact black hole solutions in shift symmetry breaking DHOST theories beyond Horndeski.  

\end{abstract}

%\pacs{04.60.Rt,04.20.Dw,04.60.-m}
%\keywords{Scalar-Tensor theories, Stealth black hole solution, Torsion}

\maketitle

Since the discovery of the acceleration of the expansion of the universe, an important effort has been devoted in constructing large scale modification of General Relativity (GR). Scalar tensor theories are by far the most studied extensions. As any modified gravity theory, these scalar-tensor candidates have to successfully explain the observed acceleration on large scale, while restoring GR on scales in which the theory is experimentally confirmed, typically for the local tests within the solar system. Additionally to the IR modifications in the cosmological sector, the new scalar-metric coupling allows to violate some of the assumptions of the famous no hair theorem which strongly restricts the black hole solution in General Relativity \cite{Volkov:2016ehx, Herdeiro:2015waa, Bekenstein:1996pn}.

%In particular, they admits black holes solutions which violate the famous no hair theorem which strongly restrict the final point of gravitational collapse in General Relativity. Owing to the famous no hair theorem, the end point of the collapse depends only on a restricted set of parameters, namely the mass $\cM$, the angular momentum $\cJ$ and the electric charge $\cQ_{\text{E}}$ measured at infinity. From this point of view, black holes are very simple objects and their classical behaviour depends only on these three parameters $(\cM, \cJ, \cQ_{\text{E}})$ which are associated to conservation Gauss laws. This property of the electro-vacuum sector of GR is summarized by the famous statement that electro-vacuum stationary black holes solutions do not support hair. See \cite{Volkov:2016ehx, Herdeiro:2015waa, Bekenstein:1996pn} for reviews.

%It is important to note that a violation of the no hair theorem could have important observational consequences. For example, from a quantum gravity perspective, the parameters $(\cM, \cJ, \cQ_{\text{E}})$ shall play the role of quantum numbers labelling the quantum black hole states. If more than three parameters are required to fully characterized black hole solutions, one could in principle miss states and a description based solely on the (quantum version of the) GR parameters $(\cM, \cJ, \cQ_{\text{E}})$ could hide degenerate states. This would be the case if quantum hairs emerged in the quantum description, as shown for example in \cite{Ghosh:2011fc}.

The search for such hairy black hole solutions beyond General Relativity is however quite challenging. Hawking and latter, Sotiriou and Faraoni, derived no hair theorems for Brans-Dicke scalar tensor theories and its generalization \cite{Sotiriou:2011dz}. Exact black hole solution with scalar hair were then found by violating some assumptions of these theorems, an example of which is the BNBB hairy black hole \cite{Bronnikov:1978mx}. However, such scalar hairy configurations are usually unstable and the scalar field fails to be regular at the horizon. An additional no hair theorem was found latter on in the more general shift symmetric Horndeski theory in \cite{Hui:2012qt}. Soon after, it was shown how to by pass the assumptions of this no-go result. Hairy solutions were obtained following two different strategies: by introducing a Gauss-Bonnet-Scalar coupling \cite{Sotiriou:2014pfa, Sotiriou:2013qea, Benkel:2016rlz, Antoniou:2017acq, Antoniou:2017hxj}, or by allowing the scalar field for a linear time-dependent profile \cite{Babichev:2013cya}. 

In this work \cite{Babichev:2013cya}, a stealth Schwarzschild-(A)dS black hole dressed with a linear time-dependent scalar field was obtained. Such stealth configuration corresponds to a vacuum metric solution of GR supplemented with a non-trivial scalar hair which does not gravitate. The scalar field is a spectator admitting a vanishing energy momentum tensor and its physical effects only show up at the perturbative level. Such stealth black hole solution were initially introduced in \cite{AyonBeato:2004ig} and represent the simplest example of hairy black hole configuration, i.e in which the scalar field remains regular on the horizon.

These stealth black hole configurations were then investigated in several scalar-tensor extensions, such as bi-scalar extension of Horndeski \cite{Charmousis:2014zaa}, covariant galileons theories \cite{Babichev:2012re}, shift symmetric beyond Horndeski \cite{Babichev:2016kdt, Babichev:2017guv, Kobayashi:2014eva, Kobayashi:2018xvr}, and more recently in shift symmetric breaking Horndeski theory \cite{Motohashi:2018wdq, Minamitsuji:2018vuw}. More details on the black holes solutions and stars within the Horndeski and beyond Horndeski classes can be found in \cite{Babichev:2016rlq, Lehebel:2018zga, Cisterna:2016vdx, Cisterna:2015yla}.

For single scalar field extension of GR, the most general theory constructed so far, up to cubic order in the derivative of the scalar field, was presented in \cite{BenAchour:2016fzp}, and dubbed degenerate higher order scalar tensor theories, i.e DHOST, owing to the degeneracy property of its lagrangian which ensures the absence of an Ostrogradsky ghost. See \cite{Langlois:2017mdk} for a recent review on DHOST theories and \cite{Langlois:2015cwa, Langlois:2015skt, Crisostomi:2016czh, Achour:2016rkg, Motohashi:2016ftl} for further details, as well as \cite{Crisostomi:2018bsp} for a recent investigation of their cosmological sector and  \cite{Crisostomi:2017lbg} concerning the Vainshtein mechanism. 
This DHOST construction encompasses most of the existing scalar-tensor candidates studied so far, among which the GLPV theory \cite{Gleyzes:2014dya, Gleyzes:2014qga}, and represents therefore an unifying framework to discuss viable scalar-tensor theories and confront them to observational tests. (See \cite{Heisenberg:2018vsk} for a review with a more general perspective). 

Recently, the joined detection of the events GW170817 and GR170817 led to the new observational constraint that the speed of gravitational wave be equal %(up to deviations of order $10^{-15}$) 
to the speed of light, i.e $c_{\text{grav}}=c_{\text{light}}$, (up to deviations of order $10^{-15}$), at least on cosmological scales \cite{GW1, GW2}.  The remaining DHOST theories satisfying this observational constraint were derived in \cite{Langlois:2017dyl}, restricting drastically the viable candidates. Additional constraint preventing from the potential decay of graviton into dark energy fluctuations were presented in \cite{Creminelli:2018xsv}.

In this letter, we provide a first scan of \textit{shift symmetry breaking} DHOST theories beyond Horndeski and derive the general conditions for a large class of these theories to admits stealth black hole configurations. We focus on hairy Schwarzschild-(A)dS and pure hairy Schwarzschild black hole solutions. Th scalar field is assumed to be linearly time-dependent while its kinetic term remains constant. 

This work is organized as follow. In section-\ref{A}, we present the general DHOST model. In section-\ref{B}, we discuss the algorithm to solve the modified field equations in the spherically symmetric case and with our scalar profile. This algorithm is borrowed from \cite{Babichev:2016kdt}. In section-\ref{C}, we present our general conditions for the hairy Schwarzschild-(A)dS and pure Schwarzschild black hole solutions. Section-\ref{D} is devoted to the subclasses satisfying the observational constraint  $c_{\text{grav}}=c_{\text{light}}$ as well as the subclass free from graviton decay. In section-\ref{E}, we consider also the reduction to the GLPV and Horndeski subclasses satisfying  $c_{\text{grav}}=c_{\text{light}}$. Finally, in section-\ref{F}, we provide concrete examples of lagrangians solutions of our conditions.

Our work extends the results obtained in previous works in several subclasses of shift symmetric DHOST theories \cite{Babichev:2017guv, Kobayashi:2014eva, Kobayashi:2018xvr, Babichev:2016kdt, Motohashi:2018wdq}  as well as on shift symmetry breaking Horndeki theory satisfying $c_{\text{grav}}=c_{\text{light}}$  \cite{Minamitsuji:2018vuw}.

%We discuss in the end how to by pass our no-go result for the GLPV subclass, using either a different hypothesis on the kinetic term or assuming matter to be minimally coupled to a different metric obtained via disformal transformation. We leave the discussion on the stability of the stealth black hole solutions reported here for future work. \smallskip

%\textbf{\textit{DHOST theories after GW170817}}

\subsection{The DHOST model}

\label{A}

Let us consider the family of DHOST theories given by the action
\bea
\label{ACTION}
 S_{\text{vDHOST}} [g, \phi] & =  \int d^4 \sqrt{|g|} \;\sum_I \cL^I \left( g, \phi \right)
\eea
where the different lagrangians read
\begin{align}
 \cL_2 & = P(\phi, X) \\
 \cL_3 & = Q \left( \phi, X\right) \Box \phi \\
 \cL_4 & = F \left( \phi, X\right) \cR \\
 \cL_5 & = A_3 \left( \phi, X\right)  \phi^{\mu} \phi^{\nu} \phi_{\mu\nu} \Box \phi+ A_4 \left( \phi, X\right) \phi^{\mu} \phi_{\lambda} \phi_{\mu\nu} \phi^{\nu\lambda} \nn \\
 & \;\;\; + A_5( \phi, X) \left( \phi_{\mu\nu} \phi^{\mu} \phi^{\nu}\right)^2 
\end{align}
where the six potentials $\left(P, Q, F, A_I \right)$ with $I \in \{ 3,4,5 \}$ are free functions of $\phi$ and its kinetic term $X$. We have adopted the notation of \cite{Langlois:2017dyl}. This family of theories corresponds to the quadratic DHOST theories amputated from the lagrangians $\cL_{1,1} =\left(\Box \phi\right)^2$ and $\cL_{1,2} = \phi_{\mu\nu} \phi^{\mu\nu}$, namely $A_1 = A_2 =0$ in the standard notation \cite{Langlois:2015cwa}.

This class of DHOST theories can be made consistent with the recent observational constraint from GW170817 which imposes that the speed of gravitons equal the speed of light (up to deviations of order $10^{-15}$), at least on cosmological scales \cite{Langlois:2017dyl}. In order to satisfy this constraint, the last two functions $A_4( \phi, X)$ and $A_5\left( \phi,  X\right)$ are related to $F( \phi, X)$ and $A_3\left( \phi,  X\right)$ through
\bea
\label{A4}
& A_4 = \frac{1}{8F} \left( 48 F^2_{X} - 8(F - X F_{X}) A_3 - X^2 A^2_3\right)  \\
\label{A5}
& A_5 = \frac{1}{2F} \left( 4F_{X} + X A_3\right) A_3
\eea
The potentials $A_3( \phi, X)$ and $F\left(\phi,   X\right)$ remain free functions, and the viable DHOST theories contain thus only four free potentials $(P, Q, F, A_3)$.

In the following, we shall derive general conditions on the potential of the DHOST family (\ref{ACTION}) to admit stealth black hole solutions without restricting ourselves to the subclass satisfying $c_{\text{grav}} = c_{\text{light}}$. This constraint, together with the constraint derived in \cite{Creminelli:2018xsv} concerning the graviton decay, will be discussed in the last section. The reduction to the beyond Horndeski (GLPV) and Horndeski theories will be also discussed at the end.
%Notice also that all the cubic terms have been ruled out by this constraint. Therefore, we are left with a scalar tensor lagrangian with at most quadratic terms in the second order scalar field derivatives. This drastically restricts the class of scalar tensor theories valid at this energy scale. 

%\smallskip
%\textit{Schwarzschild solution with linear time-dependent scalar hair}

%\textbf{\textit{Solving the fields equations: the algorithm}}

\subsection{Solving the fields equations: the algorithm}

\label{B}

In order to solve the field equations, we adopt the elegant strategy presented in \cite{Babichev:2016kdt}. Starting from the model (\ref{ACTION}), we derive the field equations that we write in a compact way
\be
\delta \cL = \cE^{(g)}_{\alpha\beta} \delta g^{\alpha\beta} + \cE^{(\phi)} \delta \phi
\ee
The field equations being rather complicated, we do not write them explicitly here. Instead, the equation of motion with respect to the metric $g_{\alpha\beta}$ can be written in the simple form
\begin{align}
\label{EOM1}
 F  G_{\alpha\beta} & = \cT_{\alpha\beta} \\
 \label{EOM2}
\left( Q_{\phi} -  Q_{X} -  P_{X}\right) \Box \phi & = \zeta  \;\;\; \; 
\end{align}
where $G_{\alpha\beta}$ is the standard Einstein tensor and where $\cT_{\alpha\beta}$ and $\zeta$ account for all the other terms obtained from the variation of the action respectively w.r.t $g_{\alpha\beta}$ and $\phi$, containing therefore all the higher order terms. 

Looking for stealth black hole solutions implies that the scalar field does not gravitate. This can be translated in (\ref{EOM1}) by $\cT_{\alpha\beta} =0$. Then, one can solve the l.h.s of the equation of motion using a GR black hole solution such that $G_{\alpha\beta}=0$. A common strategy is to assume for example a constant kinetic term for the scalar field profile, such that $X= X_{\ast}$. Then, under some specific conditions on the potentials of the Lagrangian, the effective energy momentum tensor $\cT_{\alpha\beta}$ can be written as
\be
\label{emt}
\cT_{\alpha\beta} = f(X) T_{\alpha\beta}
\ee
such that $f(X_{\ast})=0$, and example of which being $f(X) = \log{\left(X_{\ast}/X\right)}$. Below, we should derive the condition on the DHOST lagrangian (\ref{ACTION}) to admit such stealth black hole solutions.

We consider therefore a static spherical symmetric metric which reads
\bea
ds^2 = - e^{\nu(r)} dt^2 + e^{\lambda(r)} dr^2 + r^2 d\Omega^2
\eea
and we choose a linear time-dependent profile for the scalar field
\be
\label{lin}
\phi(t,r) =  \dot{\phi}_c t + \psi(r)
\ee
where $\dot{\phi}_c$ is assumed to be a constant. Following \cite{Babichev:2016kdt}, we introduce the notation $\dot{\phi}_c = M q$ which implies that the kinetic term reads
\be
X = - \frac{g^{\alpha\beta} \partial_{\beta} \phi \partial_{\alpha} \phi}{M^2} =  q^2 e^{-\nu}  - e^{-\lambda} \frac{\left(\psi'\right)^2}{M^2}
\ee
In order to further (drastically) simplify the field equations, \textit{we also assume that the kinetic term is constant everywhere, such that $X = X_{\ast} = q^2$}. Notice that the kinetic energy $-M^2X$ is negative since the gradient of the scalar field is a time-like vector. With this assumption, all the unknown potentials, commonly denoted $f(\phi, X)$, can now be written as function
\be
f \left( \phi, X\right) = f \left( qt + \psi(r), X_{\ast} \right)
\ee
In the following, we restrict further to potentials $f(\phi, X)$ satisfying
\be
\label{condition}
f_{\phi} (\phi, X_{\ast}) =0 \Rightarrow \;\; \partial^n_{\phi} f(\phi, X_{\ast}) =0 \;\;\;\; \forall n \in \mathbb{N}
\ee
This will allow us to simplify our conditions in the beyond shift symmetric case. Notice that while this condition is quite general, it is still possible to find counter-example in principle, and thus, we are potentially restricting the set of allowed potentials.

The third simplification, inherited from the assumption of a constant kinetic term, lies in that the radial dependent part of the scalar field is given by
\be
\label{psi}
\psi' = M q \sqrt{ e^{\lambda} \left( 1 + e^{-\nu} \right) }
\ee
Hence, $\psi'$ is directly known in term of the metric components, as well as its higher order derivatives: $\psi'', \psi'''$ etc. This can be plugged back in the field equations to further simplify the expression.

In the end, the field equations becomes lengthy expressions depending on the radial coordinate $r$. Now these field equations have to be satisfied at any point of space-time, and thus at any couple $(t,r)$. The elegant strategy followed in \cite{Babichev:2016kdt} is to expand the resulting field equations around a given $r_{\ast}$ and check the resulting conditions between the unknown potential $A_I(qt + \psi(r), q^2)$ and their derivatives. Denoting $\epsilon = r - r_{\ast}$, the expansion  of the equations of motion can be written as
\bea
& \cE_{tt} (r,t) = \sum^{m}_{n } \cE^{(n)}_{tt}(t,r) \big{|}_{r_{\ast}} \epsilon^{n} + \cO(\epsilon^m) = 0 \;\;\;\;\;\\
&  \cE_{rr} (r,t) = \sum^{m}_{n } \cE^{(n)}_{rr}(t,r)\big{|}_{r_{\ast}} \epsilon^{n} + \cO(\epsilon^m) = 0 \;\;\;\;\; \\
\label{ephi}
& \cE_{\phi} (r,t) = \sum^m_{n} \cE^{(n)}_{\phi} (t,r) \big{|}_{r_{\ast}} \epsilon^{n}  + \cO( \epsilon^m) = 0 \;\;\;\;\;
\eea
%Let us precise that in order to solve the field equations with this perturbative treatment, we have not used directly the e.o.m $\cE_{\phi}(t,r)$ but instead the conserved current version of this equation. Indeed, one can write instead the scalar field equation of motion as $\nabla_{\alpha} \cJ^{\alpha} = 0$ which reads
%\be
%r^2 \partial_t \cJ^t + \partial_r \left( r^2 \cJ^r\right) = 0
%\ee 
%Thanks to the linear time dependent profil, one has that $\cJ^t$ does not depend on time. This implies that $\cJ^r = C/r^2$ where $C$ is a constant of integration. Assuming that the scalar field remains regular at $r=0$ is equivalent to assume no source for this field and thus, the only choice is $C=0$, leaving us with $\cJ^r =0$. Following this argument introduced in \cite{Babichev:2016kdt}, the equation (\ref{ephi}) is actually replaced by
%\begin{align}
%\label{jr}
%\cJ_{t} (r,t)  & = \sum^m_{n =-1} \cJ^{(n)}_{t}(t,r) \big{|}_{r_{\ast}} \epsilon^{n}  + \cO( \epsilon^m) = 0 \\
%\cJ_{r} (r,t)  & = \sum^m_{n =-1} \cJ^{(n)}_{r}(t,r) \big{|}_{r_{\ast}} \epsilon^{n}  + \cO( \epsilon^m) = 0
%\end{align}
%while $\cJ^t$ is left arbitrary in the algorithm. 
The conditions we obtain out of this procedure are of the form
\begin{align}
& \cE^{(n)}_{tt}(t,r) \big{|}_{r_{\ast}} = \cE^{(n)}_{rr}(t,r)\big{|}_{r_{\ast}}  = \cE^{(n)}_{\phi} (r,t) \big{|}_{r_{\ast}} = 0 % \cJ^{(n)}_{r} (t,r) \big{|}_{r_{\ast}} = 0
\end{align}
but these conditions are not all independent. Moroever, they are only valid when evaluated at $X= X_{\ast} = q^2$. Once conditions on the potentials $\left( P, Q, F, A_I\right)$ (and their derivatives) are obtained at a given order, we inject them back in the full field equations and expand once more around the same $r_{\ast}$ to obtain new conditions. The algorithm closes when we obtain enough conditions between the potentials such that the full field equations are completely satisfied. 

Notice that this perturbative algorithm is rather general, and especially useful when working with such complicated Lagrangian. Owing to the large freedom in the potentials $A_I(\phi, X)$, the search for black hole solutions in these theories is somehow reversed, since one can start with any black hole metric and scalar profile, and using this algorithm, look for a specific Lagrangian which admits this ansatz as solution of its field equations.

Obviously, one can in principle proceed to the expansion around any value of $r$. But in practice, some specific values will allow to close the algorithm in a quicker way.
In the following, we shall expand the field equations around $r_{\ast} =0$.  The set of conditions we obtained being quite involved to reduce, we emphasize that, once the full conditions on the potentials have been found, we have checked the consistency of our solution by injecting it directly in the full (spherically symmetric reduced) field equations and check that there are identically vanishing. Having review the method of resolution of the field equations borrowed from \cite{Babichev:2016kdt}, we present now our result.

%\smallskip
%\textbf{\textit{Stealth Schwarzschild-(A)de Sitter solution with a linear time-dependent scalar field} }
\subsection{Exact hairy black hole solutions}

\label{C}

%\smallskip
We consider the Schwarzschild-(A)dS metric given by
\bea
e^{\nu} = e^{-\lambda} = 1 - \frac{2m}{r} - \Lambda r^2
\eea
The radial dependent part of the scalar field is straitforwardly obtained by integrating (\ref{psi}) and reads
\be
\psi '(r) = M q \sqrt{ \frac{ \left( 2m + \Lambda r^3 \right) r }{\left( 2m - r + \Lambda r^3 \right)^2}}
\ee
and one observes that $\psi'(r) \rightarrow 0$ when $r \rightarrow 0$. We can now inject this in the equations of motion and proceed to the expansion around $r_{\ast}=0$.
%The expansion  of the equation of motion can be written as
%\begin{align}
%\cE_{tt} (r,t) & = \sum^{m}_{n =-2} \cE^{(n)}_{tt}(t) \; r^{n} + \cO(r^m) = 0 \\
% \cE_{rr} (r,t) & = \sum^{m}_{n = - 1/2} \cE^{(n)}_{rr}(t) \; r^{n} + \cO(r^m) = 0 \\
%\cJ_{t} (r,t)  & = \sum^m_{n =-1} \cJ^{(n)}_{t}(t)  r^{n}  + \cO( r^m) = 0 \\
%\cJ_{r} (r,t)  & = \sum^m_{n =-1} \cJ^{(n)}_{r}(t) r^{n}  + \cO( r^m) = 0
%\end{align}

\subsubsection{Stealth Schwarzschild-(A)dS solution}
Applying the algorithm reviewed above, we obtain a complete set of conditions on the potentials which fully solve the field's equations. These conditions, valid only when evaluated at the value $X_{\ast} =  q^2$, read
\begin{align}
\label{c} 
 Q_{\phi} \big{|}_{X_{\ast}} = F_{\phi}  \big{|}_{X_{\ast}} & =  0 \\
  \label{co}
 Q_{X}\big{|}_{X_{\ast}} & = - \frac{X}{2} A_{3\phi}\big{|}_{X_{\ast}} \\
 \label{cc} 
 P \big{|}_{X_{\ast}} & = - 6 \Lambda F \big{|}_{X_{\ast}} \\
 \label{ccc} 
A_3 \big{|}_{X_{\ast}} & =  \frac{2}{9 X \Lambda} \left( P_{X} + 12 \Lambda F_{X}\right) \big{|}_{X_{\ast}}
\end{align}
%Notice that one can have $F_{\phi} \big{|}_{X_{\ast}} =0$ and at the same time,  $F_{X} \big{|}_{X_{\ast}} \neq 0$ and $F_{\phi X} \big{|}_{X_{\ast}}\neq 0$, as shown for example by
%\be
%F(\phi, X) = f(\phi) h (X)
%\ee
%where $f(\phi)$ and $h(X)$ are two functions such that $h(X_{\ast})=0$ but $h_X(X_{\ast}) \neq 0$. 
At this stage, there are no condition on the potentials $A_4$ and $A_5$, and the class solution of our conditions depend still on sic free potentials $(P, Q, F, A_I)$.
%Moreover, we see that from (\ref{co}) and (\ref{ccc}) that $A_{3\phi} = - 2 Q_X/X$. `
With this conditions (\ref{c}) to (\ref{ccc}), we have obtained a subset of DHOST theories, larger than the sector staisfying $c_{\text{grav}} = c_{\text{light}}$, which admits a stealth Schwarzschild-(A)dS solution dressed with a linear time dependent scalar field (\ref{lin}), assuming a constant kinetic term $X= X_{\ast} =  q^2$.
%The DHOST subclass of \cite{Langlois:2017dyl}, solution of our algorithm, is given by
%\begin{align}
%\cL[g, \phi] & = F(\phi, X) \left(  \cR - 6 \Lambda \right) + Q(\phi, X) \Box \phi  \nn  \\
%& \;\;\;- A_3 (\phi, X)\phi^{\mu} \phi^{\nu} \phi_{\mu\nu} \Box \phi  + A_4(\phi, X) \phi^{\mu} \phi_{\lambda} \phi_{\mu\nu} \phi^{\nu\lambda}  \nn \\
%\label{lag}
%& \;\;\; + A_5(\phi, X)  \left( \phi_{\mu\nu} \phi^{\mu} \phi^{\nu}\right)^2 
%\end{align}

%\begin{align}
%\label{c22}
%A_4  = - A_3 \left(  1 + X^2 \frac{ A_3}{8 A_2} \right)  \qquad A_5  = \frac{X A^2_3}{2 A_2} 
%\end{align}
%Plugging these conditions, we obtain the following lagrangian
%\begin{align}
%\cL[g, \phi] & = \zeta  \cR + A_3(X) \phi^{\mu} \phi^{\nu} \phi_{\mu\nu} \Box \phi  \nn  \\
%& \;\;\;  - A_3(X) \left(  1 + X^2 \frac{ A_3(X)}{8 A_2} \right)  \phi^{\mu} \phi_{\lambda} \phi_{\mu\nu} \phi^{\nu\lambda}  \nn \\
%\label{lag}
%%& \;\;\; + \frac{X A^2_3(X)}{2 A_2}   \left( \phi_{\mu\nu} \phi^{\mu} \phi^{\nu}\right)^2 
%\end{align} 

%{ \Jib Additionally, running again the algorithm for the static case which corresponds to $q=0$, one obtains only condition (\ref{cc}). }

\subsubsection{Stealth Schwarzschild solution: $\Lambda = 0$}

It is interesting to investigate the case $\Lambda = 0$ which corresponds to a pure Schwarzschild geometry. In that case, several of the previous conditions are modified because some of them are proportional to $\Lambda$ and therefore disappear for pure Schwarzschild. The new set of conditions, valid at $X = X_{\ast}$, are
\begin{align}
\label{o} 
  Q_{\phi} \big{|}_{X_{\ast}}= F_{\phi} \big{|}_{X_{\ast}} & =  0 \\
  \label{ooo}
   P\big{|}_{X_{\ast}} =P_{X} \big{|}_{X_{\ast}} & = 0\\
 \label{oo} 
 Q_{X}\big{|}_{X_{\ast}} & = - \frac{X}{2} A_{3\phi}\big{|}_{X_{\ast}}
\end{align}
Once again, there are no condition on the potentials $A_4$ and $A_5$ which remain free. In total, this class depends still six free potentials $(P, Q, F, A_I)$ subject to conditions (\ref{o})-(\ref{oo}) at $X= X_{\ast}$.

We can now investigate additional requirement for the subclass of DHOST that we found in order to satisfy the recent observational constraints. % beyond-Horndeski subclass, i.e the GLPV theory \cite{Gleyzes:2014dya, Gleyzes:2014qga}, to admit the stealth Schwarzschild-(A)dS or pure stealth Schwarzschild solution with our scalar profile while accounting for $c_{\text{grav}}=c_{\text{light}}$. %This will provide a comparison with previous works, among which \cite{Babichev:2016kdt}. 

%As a first step, let us rewrite the quartic beyond Horndeski (or GLPV) lagrangian, in the form
%\begin{align}
%\cL_4^{H} + \cL_4^{bH} & = G_4(\phi, X) R \nn \\
%& - X \left( F_4(\phi, X) + \frac{2G_{4X}(\phi, X)}{X}\right) \left( \left(\Box \phi\right)^2 - \phi_{\mu\nu} \phi^{\mu\nu}\right) \nn \\
%& + 2F_4(\phi, X) \phi_{\mu\nu} \phi^{\mu} \phi^{\nu} \Box \phi - 2 F_4(\phi, X)  \phi^{\mu} \phi_{\lambda} \phi_{\mu\nu} \phi^{\nu\lambda} 
%\end{align}
%where we have re-introduced the standard notation, such that $A_2 = G_4$ and $A_3 = 2 F_4(X)$. 
\subsection{Observational constraints}

\label{D}

\subsubsection{Subsector satisfying $c_{\text{grav}} = c_{\text{light}}$ }

We can now impose the recent observational constraint $c_{\text{grav}} = c_{\text{light}}$ and reduce the above constraints to the viable subsector of DHOST remaining after GW170817 \cite{Langlois:2017dyl}. 

\textit{Schwarzschild-(A)dS:} This is done by imposing the relations (\ref{A4}) and (\ref{A5}) to obtain the two last potentials $A_4$ and $A_5$ at $X= X_{\ast}$. These conditions read for $A_4(\phi, X)$
\bea
\label{A44}
 & A_4  \big{|}_{X_{\ast}} =  \frac{6 F_X^2}{F} -\frac{\left(12 \Lambda  F_X+P_X\right)^2}{162 \Lambda ^2 F_X}+\frac{2 (X-1) \left(12 \Lambda  F_X+P_X\right)}{9 \Lambda  X} \big{|}_{X_{\ast}} \;\;\;\;\; 
\eea
while for $A_5(\phi, X)$, one has
\bea
\label{A55}
 A_5 \big{|}_{X_{\ast}} = \frac{2 \left(12 \Lambda  F_X+P_X\right) \left(30 \Lambda  F_X+P_X\right)}{81 \Lambda ^2 X F} \big{|}_{X_{\ast}} \;
\eea
which ensure that $c_{\text{grav}} = c_{\text{light}}$ for the Schwarzschild-(A)dS solution sector. 

\textit{ Pure Schwarzschild:} In that case, we don't have any constraint on $(A_3, A_{3X}, F, F_{X})$ at $X= X_{\ast}$, there is no additional constraint on $A_4$ and $A_5$ for this subsector.
Notice that despite the above constraints on $A_4$ and $A_5$ at $X= X_{\ast}$, the subset of theories satisfying (\ref{A44}) and (\ref{A55}) has again six free potentials $(P, Q, F, A_I)$.

\subsubsection{Absence of graviton decay}

If one takes into account now the recent constraint derived in \cite{Creminelli:2018xsv} for the absence of graviton decay, in addition to the constraint $c_{\text{grav}} = c_{\text{light}}$, then the DHOST action (\ref{ACTION}) reduced to
\begin{align}
\cL &= P(\phi, X) + Q(\phi, X) \Box \phi + F(\phi, X) R  \nn \\
& + 6 F^2_X(\phi, X) / F(\phi, X) \;   \phi^{\mu} \phi_{\lambda} \phi_{\mu\nu} \phi^{\nu\lambda}
\end{align}
with only three free potentials $(P, Q, F)$. It implies that 
\be
A_3(\phi, X) =0 \qquad A_4(\phi, X) = 6 F^2_X(\phi, X) / F(\phi, X) 
\ee
\textit{Schwarzschild-(A)dS:} The constraint (\ref{ccc}) implies that
\be
F_X \big{|}_{X_{\ast}} = - \frac{P_X }{12\Lambda} \big{|}_{X_{\ast}}
\ee
which, as expected, automatically satisfies the constraint on $A_4$ at $X= X_{\ast}$ as seen from (\ref{A44}).

\textit{Pure Schwarzschild:} In that case, the vanishing of $A_3(\phi, X)$ implies from (\ref{oo})
\be
Q_X\big{|}_{X_{\ast}} =0
\ee
and no additional restriction than (\ref{o})-(\ref{ooo}) occurs. 

It implies therefore that the subclass of shift symmetry breaking DHOST theories successfully accounting for the observational constraints $c_{\text{grav}} = c_{\text{light}}$, and satisfying the requirement of absence of graviton decay, still admits non trivial stealth black hole solutions. In each case, the remaining theory has still three free potentials $(P, Q, F)$ still depending on $\phi$ and $X$.

\subsection{Beyond Horndeski and Horndeski subclasses}

\label{E}

We consider now the GLPV and Horndeski subclasses satisfying $c_{\text{grav}} = c_{\text{light}}$. These two subclasses are given, in the standard notation, by $P = G_2$, $Q = G_3$ and  $F = G_4$ with
\begin{align}
\label{GLPV}
 A_3 & = - A_4 = -4F_X/X \qquad\text{GLPV} \\
 A_3 & = A_4 = F_X =0   \qquad\text{Horndeski}
\end{align}
 and $A_5 =0$ valid for any $\phi$ and $X$. The resulting lagrangian depends on only three free potential $(P, Q, F)$.
  
%Imposing $A_5 =0$ from (\ref{A5}), and using (\ref{GLPV}), one has two possibilities, namely
%\begin{align}
%\label{case2}
%A_3  & = - 4 F_X /X = - 4 F/X \qquad \;\; \text{GLPV} \\
%\label{case1}
%A_3 & = F_X = 0\qquad  \qquad \qquad  \qquad\text{Horndeski}
%\end{align}

\textit{Schwarzschild-(A)dS:} Using (\ref{A44})-(\ref{A55}), one obtains the following constraints in each cases
\begin{align}
\label{co}
 F_X \big{|}_{X_{\ast}} &  =  - \frac{P_X}{30\Lambda} \big{|}_{X_{\ast}}   \qquad \;  \text{GLPV} \\
 \label{coo}
{P_X}  \big{|}_{X_{\ast}} =  {Q_X}  \big{|}_{X_{\ast}} & =0 \qquad\qquad \;\;\; \text{Horndeski}
\end{align}
additionally to the constraint (\ref{c}) and (\ref{cc}).

\textit{Pure Schwarzschild:} For this solution, one obtains from (\ref{oo}) the following constraints
\begin{align}
\label{cooo}
 %Q_{X}\big{|}_{X_{\ast}}  = - \frac{X}{2} A_{3\phi} \big{|}_{X_{\ast}}  & =  2 F_{X\phi}  \big{|}_{X_{\ast}}
\text{No additional constraints}&\;\;\;\; \qquad  \qquad \text{GLPV}  \\
 \label{coooo}
  Q_{X}\big{|}_{X_{\ast}}   =0 & \qquad \qquad  \text{Horndeski} 
 \end{align}
 additionally to constraints (\ref{o})-(\ref{ooo}). Notice that in the Horndeski case, since $F$ depends only on $\phi$, constraint (\ref{o}) implies that $F(\phi) =\zeta$ where $\zeta$ is a constant
 \footnote{Notice that this last result is not in contradiction with the no-go result concerning the existence of stealth black hole solutions with time-dependent scalar profile obtained in shift symmetry breaking Horndeski theory with $c_{\text{grav}} = c_{\text{light}}$ in \cite{Minamitsuji:2018vuw}. Indeed, this result assumes $\partial_{\mu}X \neq 0$.}.
% \begin{align}
%Q_X \big{|}_{X_{\ast}}  & =0 \qquad\qquad \;\;\; \text{Horndeski}
%\end{align}
%additionally to constraints (\ref{o})-(\ref{ooo}).

\subsection{Concrete examples}

\label{F}

As a last step, we provide concrete examples of potentials solutions of our conditions for both the Schwarzschild-(A)dS and pure Schwarzschild cases for the general DHOST case, but also for the GLPV and Horndeski subclasses.

\subsubsection{Examples for DHOST}

Let us first focus on the general DHOST lagrangian. 

\textit{Schwarzschild-(A)dS:} An example of solution of our conditions (\ref{c})-(\ref{ccc}) is given by
\begin{align}
	& F (\phi, X )= f_1(\phi, X)  \log{\left(\frac{q^2}{X}\right)} + f_2(X)\\
	& P(\phi, X ) = f_3(\phi, X)  \log{\left(\frac{q^2}{X}\right)} -6 \Lambda  f_2(X)\\
	& A_3(\phi, X ) = \frac{2}{9 X \Lambda} \left[ 6 \Lambda f_{2X}(X) -   \frac{f_{3}(\phi, X)  + 12 \Lambda f_1(\phi, X)  }{X}  \right]  \\
	& Q(\phi, X ) = - \frac{1}{9 \Lambda} \left[ f_{3 \phi}(\phi, X)  + 12 \Lambda f_{1 \phi}(\phi, X)  \right] \log{\left(\frac{q^2}{X}\right)} \notag \\
	& \qquad \qquad + \int dX \log{\left(\frac{q^2}{X}\right)} f_4(X)
\end{align}
where $(f_1,f_3)$ are free potentials depending on both $\phi$ and $X$ while $(f_2,f_4)$ are free potentials depending only on $X$.

\textit{Pure Schwarzschild:} For the pure Schwarzschild solution, an example of potentials solving our conditions (\ref{o})-(\ref{oo}) is given by
\begin{align}
	& F (\phi, X ) = f_1(\phi, X)  \log{\left(\frac{q^2}{X}\right)} + f_2(X)\\
	& P(\phi, X ) = f_3(\phi, X)  \log^2{\left(\frac{q^2}{X}\right)} \\
	& Q(\phi, X ) =  \frac{X^2}{2} A_{3 \phi} (\phi, X) \log{\left(\frac{q^2}{X}\right)} \notag \\
	& \qquad \qquad \qquad \qquad \qquad + \int dX \log{\left(\frac{q^2}{X}\right)} f_4(X)
\end{align}
where again, $(f_1,f_3)$ are free potentials depending on both $\phi$ and $X$ while $(f_2,f_4)$ are free potentials depending only on $X$.

\subsubsection{Examples for GLPV}

We focus now on the GLPV subclass satisfying the observational constraint $c_{\text{grav}} = c_{\text{light}}$.

\textit{Schwarzschild-(A)dS:} A set of potentials solutions of our conditions is given for example by
\begin{align}
	& F (\phi, X )= f_1(\phi, X)  \log{\left(\frac{q^2}{X}\right)} + f_2(X)\\
	& P(\phi, X ) & \notag \\
	& =  \left[- 30 \Lambda f_1(\phi, X) + 24 \Lambda   X f_{2X}(X) \right] \log{\left(\frac{q^2}{X}\right)}   -6 \Lambda  f_2(X)    \\
	& A_3(\phi, X )& \notag \\
	&  = - \frac{4}{X} \left[ f_{1X}(\phi, X)  \log{\left(\frac{q^2}{X}\right)} - \frac{ f_{1}(\phi, X)}{X}  + f_{2X}(X) \right]\\
	& Q(\phi, X ) = 2 \left[ f_{1 \phi}(\phi, X)  \right] \log{\left(\frac{q^2}{X}\right)}  +  \int dX \log{\left(\frac{q^2}{X}\right)} f_{4}(X)
\end{align}
which depends again on four free potentials.

\textit{Pure Schwarzschild solution:} For the pure Schwarzschild case, an example of potentials solutions of our conditions is given by
\begin{align}
	& F (\phi, X ) = f_1(\phi, X)  \log{\left(\frac{q^2}{X}\right)} + f_2(X)\\
	& P(\phi, X ) = f_3(\phi, X)  \log^2{\left(\frac{q^2}{X}\right)} \\
	& A_3(\phi, X ) = - \frac{4}{X}  (f_{1X}(\phi, X)  \log{\left(\frac{q^2}{X}\right)} - \frac{ f_{1}(\phi, X)}{X} + f_{2X}(X))\\
	& Q(\phi, X ) = 2  f_{1 \phi}(\phi, X)  \log{\left(\frac{q^2}{X}\right)}  +  \int dX \log{\left(\frac{q^2}{X}\right)} f_{4}( X)
\end{align}

\subsubsection{Examples for Horndeski}

Finally, let us now consider the Horndeski theory with $c_{\text{grav}} = c_{\text{light}}$. An example of Horndeski lagrangian admitting a stealth Schwarzschild-(A)dS black hole solution for our specific scalar profile is the standard Einstein-Hilbert term plus a cosmological constant together with a generalized Galileon term, given by
\begin{align}
\cL_{\text{Horndeski}} = \cR -  6\Lambda + \log^2{\left(\frac{q^2}{X}\right)} q(\phi, X) \Box \phi 
\end{align}
where $q(\phi, X)$ is a free potential. One can easily check that it satisfies conditions (\ref{coo}). The pure stealth Schwarzschild solution is simply obtained for $\Lambda=0$.
As mentioned in the introduction, the square log term allows to obtain an effective energy momentum tensor for the scalar field of the form (\ref{emt}), which vanishes for $X=q^2$. Therefore, the scalar field does not gravitate and the metric remains a pure GR solution. This provides a simple example of the mechanism behind stealth black hole configuration in such higher order scalar-tensor theories.

%Therefore, the subclass of the GLPV theory statisfying $c_{\text{grav}} = c_{\text{light}}$ does not admit our stealth Schwarzschild-(A)dS solution with a linear time-dependent profil and a constant kinetic term. However, the shift symmetric GLPV subclass admits a stealth Schwarzschild solution, at least for the specific scalar profil ansatz (\ref{lin}), compared with the Horndeski subclass satisfying $c_{grav}=c_{light}$, where, as recently shown in \cite{Minamitsuji:2018vuw}, stealth solution with the specific scalar profil ansatz (\ref{lin}) does not exist. Moreover, the conditions (\ref{o}-\ref{oo}) for a stealth Schwarzschild solution provide a generalization of the stealth solutions found previously in \cite{Kobayashi:2014eva, Kobayashi:2018xvr} for the shift symmetric GLPV and DHOST theories. \\

\section{Conclusion}

In this letter, we have provided a first scan of the spherically symmetric sector of a large family of \textit{shift symmetry breaking} DHOST theories (the quadratic lagrangian with $A_1=A_2=0$). We have derived general conditions for this subclass to admit the vacuum Schwarzschild-(A)dS solution, Eq.(\ref{c})-(\ref{ccc}), or pure Schwarzschild solution of GR, Eq.(\ref{o})-(\ref{oo}), with a linear time-dependent scalar dressing as well as for a static scalar dressing under the assumption of a constant kinetic term, i.e $X = q^2$. 

Then, we have restricted the class of theories by requiring the remaining shift symmetric DHOST theories, solutions of our conditions, to satisfy the observational constraint $c_{\text{grav}}=c_{\text{light}}$ as discussed in \cite{Langlois:2017dyl}, and finally, the recent theoretical requirement of absence of graviton decay discussed in \cite{Creminelli:2018xsv}. For all these viable subclasses, we have shown that stealth black hole solutions, with our without a cosmological constant, exist. 

Finally, we have consider the restriction to the shift symmetry breaking GLPV and Horndeski subclasses, statisfying the constraint $c_{\text{grav}}=c_{\text{light}}$. We have shown that in each subclass, one can also find stealth black holes solutions upon satisfying the conditions (\ref{co})-(\ref{coooo}). In the last section, we have provided concrete examples satisfying the conditions we found.

This work extends previous results focusing on stealth  black hole solution in DHOST with a constant scalar profile \cite{Motohashi:2018wdq}, in shift symmetry breaking Horndeski theory with $c_{\text{grav}}=c_{\text{light}}$ for more general scalar profiles \cite{Minamitsuji:2018vuw}, as well as previous results obtained for linear time-dependent scalar dressing in shift symmetric GLPV and DHOST theories with or without $c_{\text{grav}}=c_{\text{light}}$ \cite{Babichev:2017guv, Kobayashi:2014eva, Kobayashi:2018xvr, Babichev:2016kdt}.

A crucial step for the future would be to investigate the fate of the stealth black hole solutions considered here at the perturbative level to fully contemplate the scalar field back-reaction on the metric and investigate its stability. Indeed, it is well known that the stealth hairy black hole solutions found in shift symmetric Horndeski theory with $X= \text{Constant}$, are either unstable or strongly coupled as shown in \cite{Ogawa:2015pea}. We leave this stability analysis for future work.

\bigskip \textit{Acknowledgement}: We are grateful to H. Motohashi and M. Minamitsuji for useful discussions and for having pointed to us a mistake in the shift symmetry limit of our conditions in the previous version of this draft, thank to \cite{Motohashi:2019sen}. We would like also to thanks C. Charmousis, D. Langlois and K. Noui for their comments on a first version of this work. This work was supported by the National Science Foundation of China, Grant No.11475023 and No.11875006 (J. BA), as well as by the China Postdoctoral Science Foundation  with Grant No. 212400209 (J. BA).

\end{document}